\documentstyle[sprocl]{article}

\begin{document}

\title{Rochelle salt: a prototype of particle 
physics\footnote{Plenary talk given at the International Workshop COSMO97
on Particle Physics and the Early Universe, 15-19 September 1997,
Ambleside, Lake District, England.}}
\author{ Goran Senjanovi\'c}
\address{{\it International 
Center for Theoretical Physics, 34100 Trieste, Italy 
}}
\maketitle
\begin{abstract}
Rochelle salt has a remarkable characteristic of 
becoming
more ordered for a range of high temperatures before 
melting. In the particle physics language
this means more symmetry breaking for high T. In many 
realistic field theories this is a perfectly  
consistent
scenario which  has profound consequences in the 
early 
universe. In particular it implies that there may be 
no domain wall and monopole problems, and it may also 
play  an important role in baryogenesis if CP and P are 
broken spontaneously. In the case of the monopole problem 
this may require a large background charge of the 
universe.
The natural candidates for this background charge are a 
possible lepton number in the neutrino sea or global 
continous R-charges in supersymmetric 
theories.

\end{abstract}

\section{Introduction}

Intuition and experience tell us that with increasing 
temperature physical systems become less ordered. It is 
appealing to believe that this is a universal 
physical
law, but surprisingly enough there are exceptions. 
The well known counterexample is a Rochelle salt which, 
when heated up crystalizes more, at least for a range of 
temperatures, until it eventually melts\cite{jona-shirane}. 
This is a remarkable phenomenon and one would like to 
know how general it is. It turns out to be a natural 
possibility in many realistic particle physics theories. We can 
divide its  source in two different categories:

a) {\it microscopic properties of the theory}. Here 
we have 
a range of parameters in the Lagrangian which allows 
for 
nonvanishing vevs at high temperature\cite{w74,ms79a,ms79b,ms79c}. 
The theory in question must have at 
least two Higgs  multiplets, a natural
feature of any extension of the standard model (SM).   

b) {\it macroscopic external conditions}. It is 
exemplified 
by a large background charge density of the universe 
and has nothing to do with the underlying  
microscopic theory. In 
this case, for large enough charge density, 
symmetries get 
broken at high T in all of the parameter space 
\cite{l76,l79}. The natural
candidates for such charges are lepton number in SM 
\cite{l79,ls94,brs97} and
continous global R-charges in supersymmetry 
\cite{rs97,bdrs98}.

  Both scenarios are very appealing; however, they are 
far 
from being automatic. The questions are:

i) why should we live in the parameter space that 
allows
for high T symmetry breaking in  case a)? ;
and
ii) what could have created large background charge 
in  case b)?
 We have no answers yet to these questions. However, 
the consequences of this phenomenon are striking, and
worth discussing. First of all, if at high T 
symmetries do not get restored, there may be no domain wall 
\cite{ms79b,ds95}and
monopole problems \cite{sss85,dms95}. It has been known for 
a long time that during phase transitions from the unbroken 
to the broken phase topological defects get 
formed \cite{k76}. In the
case of discrete symmetries the resulting defects, 
the domain walls, are a cosmological catastrophy,  
since a single 
large wall carries far too much energy density 
\cite{zko74}. The 
monopole problem is rather different: a single 
monopole
poses no problem at all, but during the GUT phase 
transition  we get too many of them \cite{p79}. If, 
on the other hand, there 
is no phase transition, these problems would simply 
dissapear. This is similiar to inflation, and   
should be not viewed as an alternative to it, but 
rather as a complementary phenomenon. As we  discuss 
below, inflation 
better take place (after all, it is the solution to 
the horizon problem). However, it does not have to  
take place at lower temperatures. This may be of great help in 
model building.

    Second, symmetry breaking at high T may play an 
important role in baryogenesis, if it takes  place at 
temperatures much above the weak scale and if CP and 
P are spontaneously broken at lower scales. 
Spontaneous breaking of P and T symmetries provides an 
alternative to 
the axion as the solution of the strong CP problem 
\cite{bt78,ms78,g78}. Thus, 
if symmetries are not restored at high temperature,
parity and time-reversal symmetries would 
remain broken  as to allow for a 
nonvanishing baryon density \cite{ms80b}.

\section{Symmetry breaking at high temperature}

We now discuss the possibility of our particle 
theories 
mimicking Rochelle salt. We wish to achieve symmetry 
breaking at high T, i.e. we wish to have a scalar field 
$\phi$ possess a nonvanishing VEV for $T\gg m$, where m 
is the relevant physical scale. Since in such a case T 
becomes the only scale of the theory, one expects for 
$T\gg m$

\begin{equation}
\langle \phi (T) \rangle \simeq T
\end{equation}

and thus we would have more order with incrasing 
temperature. In other words the effective mass   term
for $\phi$ at high T needs to be negative.
We have  already said that the sources of this may be 
either microscopic or macroscopic, which we now 
discuss.

\subsection{Microscopic}

This is purely a property of an underlying theory and 
it
requires at least two scalar multiplets. Namely, the 
effective  mass term for a field $\phi$ at high T 
has the form \cite{w74}

\begin{equation}
\mu^2 (T) = (g^2 + |h|^2 + \lambda) T^2
\end{equation}
 where $g$, $|h|$ and $\lambda$ stand for the gauge, 
Yukawa
and scalar contribution, respectively. Also, the 
equation 
is symbolic in a sense that the precise coefficients 
are 
ommitted since they play no role in the qualitative 
picture
we are discussing here.
 The first two terms
are manifestly positive, and in the single field case 
so is
the last one, since $\lambda >0$ is the necessary 
condition
for the boundedness of the potential. However, if 
there are
more scalar fields, some of their couplings are 
allowed to be  negative and the $\mu^2 (T)$ need not 
be necessarily 
positive. There is a finite parameter space which 
corresponds  to a negative high T mass term and a 
nonvanishing VEV \cite{w74,ms79a,ms79b}. 
However, for realistic values of gauge couplings this 
parameter space becomes  be very small, and next to 
the leading terms seem to invalidate this picture 
\cite{bl96b}. This is
a crucial fact to keep in mind when we discuss the 
monopole problem below. On the other hand, in the case
of global symmetries without large Yukawa couplings,  
the high T symmetry nonrestoration is a perfectly valid
scenario \cite{r96,a-c96,o97,prt97} and it plays an 
important role for the domain wall problem. 

What happens in supersymmetric theories? The learned 
reader could have already noticed that the above mechanism 
of symmetry nonrestoration  is not compatible
with supersymmetry \cite{h82,m84}, since supersymmetry relates 
Yukawa and scalar couplings and we have already argued that 
Yukawa contribution to the mass term is always positive.
It does not help to include the nonrenormalizable 
interactions \cite{bms96,bs96}, although there has been 
some  
promise 
originally \cite{dt96}. On the other hand, in 
theories with flat directions the idea of 
nonrestoration seems to  work
\cite{dkl97}.

\subsection{Macroscopic}

This is the case of the nonvanishing background 
charge 
density of the universe. To illustrate the 
phenomenon,
take a simple case of the complex scalar field $\phi$ 
with a global $U(1)$ symmetry  

\begin{equation}
\phi \to e^{i\alpha} \phi
\end{equation}

and let us assume a nonvanishing background charge 
density
$n$ corresponding to the charge $Q$: $Q=n V$ ,where V 
is the volume. Notice that if the charge $Q$ is 
conserved during 
the expansion of the universe, the density grows with 
the
temperature: $n \simeq T^3$, since $V \simeq T^{-3}$. 
The effective potential for the 
field $\phi$ at high T ($T \gg m$, where m is the 
$T=0$
mass term) and high density $n \simeq T^3$ is readily 
found to be \cite{hw82,bbd91} 

\begin{equation}
V(n,t) = {n^2 \over 2(|\phi|^2 + T^2/3)} + {\lambda 
\over 6} T^2 |\phi|^2  + {\lambda \over 4} |\phi|^4
\end{equation}

It is clear that the first term prefers $\phi$ to be 
nonvanishing,  as opposed to the second high T mass 
term. 
For sufficiently large density n

\begin{equation} 
n > n_C = {1 \over 3} \sqrt{\lambda \over 3 } T^3
\end{equation}

$\phi$ has a nonvanishing VEV
and the symmetry is broken independently of what 
happens
at $T=0$ \cite{bbd91}. There is nothing mysterious 
about 
symmetry breaking in this case: for sufficiently 
large
density it becomes more advanageous for the system to 
store 
the charge in the vacuum rather than in the thermal 
modes.
This is what we meant by macroscopic: symmetry
breaking at high T is due to the macroscopic 
conditions
in the universe and has nothing to do with the space 
of
the parameters of the microscopic theory. All that is 
needed  is a sufficiently large background charge on 
the order of
the entropy (actually even smaller), a rather natural
condition. Of course, whether or not it is easy to 
achieve
this condition is not so clear and  requires more 
serious
study. 

  A more interesting question for us is  what 
charge can play this role. In the standard model we have a 
perfect candidate:  lepton number. At low T, lepton number 
is a perfect symmetry, at least on cosmological time 
scales, and we have a neutrino sea in the universe with a density 
on the order of the photon density, i.e. on the order of 
entropy. If
the neutrino sea were to carry a lepton number, the 
gauge
symmetry of the SM would remain broken at high T. We 
shall
discuss the consequences in the following section.

An important feature of this phenomenon is that it is 
equally operative in supersymmetric theories 
\cite{rs97}. I discuss it here from the conceptual point
of view; for computational and technical aspects see
the talk of Borut Bajc at this conference \cite{borut}. 
Actually in supersymmetry there is another perfect 
candidate
for the background charge. Many supersymmetric models
possess global continous R-charges, i.e. charges that do 
not 
commute with supersymmetry. In fact, in the 
supersymmetric
standard model there is an automatic $U(1)$ R-symmetry,
even if one allows all the gauge invariant terms in 
the superpotential, including those that break matter 
parity (R-parity). If the universe had a large background R-
charge in the early universe, the gauge symmetries of  the MSSM 
would have been broken at high T. Of course, soft 
supersymmetry
breaking terms also break  the continous R-symmetry 
and thus eventually wash out the original R-charge. It is easy 
to estimate the rate of the R- breaking processes to be 
of order \cite{bdrs98}

\begin{equation}
\Gamma_R \simeq \sqrt{m_S T}
\end{equation}

where $m_S \simeq m_W$. Thus for temperatures below 
$10^7-10^8 GeV$
the R-breaking processes are in thermal equilibrium
and they will wash out any memory of the previous 
charge.
This is a remarkable situation. We may have a 
dramatic
impact of R-charges on cosmology  without them leaving any 
trace today.
 
  It is interesting to see what happens in the 
context of GUTs.  In general R-symmetries are not 
automatic and, in fact,
in the minimal supersymmetric $SU(5)$ GUT there is no 
such 
symmetry. On the other hand, in the minimal model the 
GUT 
scale is put in by hand, or better yet, the ratio 
between the  GUT and the weak scale is fine-tuned. It 
is far more appealing to  have this ratio determined 
dynamically 
through radiative corrections and soft supersymmetry
breaking. Fortunately, this attractive scenario cries
for R-symmetry. 

Let me illustrate this on a simple
model \cite{bdrs98} based on $SU(6)$ grand unified 
theory and an adjoint
representation superfield $\Phi$. If one adopts a 
philosphy of not 
introducing any mass terms by hand, the most general 
superpotential for $\Phi$ has the form 

\begin{equation}
W=\lambda Tr \Phi^3
\end{equation}

It has a manifest $U(1)$ R-symmetry $\Phi \to e^{i 
\alpha} \Phi$, $\theta \to e^{3i\alpha/2} \theta$, i.e.  
$\phi \to e^{i\alpha} \phi$, $ \psi \to e^{-i\alpha/2} 
\psi$,
where $\phi$ and $\psi$ are the scalar and fermionic 
components of the superfield $\Phi$. 
At zero temperature, from
\begin{equation}
F_{\phi} = \lambda (\phi^2 - {1 \over 6} Tr \phi^2)=0
\end{equation}
 
and with $\phi$ diagonal as to make $V_D$ vanish, the 
supersymmetric minimum has a flat direction

\begin{equation}
\phi = \phi_0 diag(1,1,1,-1,-1,-1)
\end{equation}
  The flat direction is a result of the 
original
R-symmetry and it will be lifted by the soft 
supersymmetry 
breaking terms. In the usual manner the scale 
$\phi_0$ is
then determined radiatively and it is naturally 
superlarge. 
The original $SU(6)$ symmetry gets broken to its 
$SU(3) \times SU(3) \times U(1)$ subgroup, which 
implies
the existence of monopoles.
We shall not dwell on this here, for us it is 
sufficient to 
note the role that R-symmetry plays in this picture.
In the same manner as before, if the background 
charge 
is large enough, the $SU(6)$ symmetry will not be 
restored
at temperatures above the GUT scale. 
The effective potential at high temperature and high
density has a form

\begin{equation}
V(n,T)={n^2 \over 2 (4Tr\Phi^\dagger \Phi  + 
 (105/4)\,T^2 )}
+6 g^2 T^2 Tr\Phi^\dagger \Phi + V_F + V_D
\end{equation}
where $V_F$ and $V_D$ are the $T=0$ potentials. It is 
easy to see that in this case for the density bigger 
than the
critical one
\begin{equation}
n > n_C= {105 \over 4} \sqrt{3} g T^3
\end{equation}
the symmetry remains broken and no phase transition 
takes
place.
This provides a solution to the monopole problem. 
One can implement the same idea in the $SU(5)$ 
theory,
but in order to make it work one needs to increase 
the Higgs
sector to two adjoint and one singlet representation
\cite{bdrs98}. The model is identical in spirit to 
 Witten's original idea \cite{w81}.

\section{Discussion and outlook}

We have seen above that the idea of high T symmetry 
breaking  is quite legitimate and has important 
cosmological 
consequences. Let us discuss the most important ones.

\subsection{Domain wall problem}

The phenomenon of symmetry nonrestoration  in 
general 
works perfectly well, as long as the discrete 
symmetry in
question is broken by a gauge singlet field. There 
are numerous examples of singlets in this role, the  
most notable 
one being the invisible axion model which suffers 
from 
the axionic domain walls \cite{s82}. They get formed 
at the temperature  on the order of the QCD phase 
transition when the walls
get attached to strings formed earlier \cite{ve82}. 
All we need is 
to eliminate the original phase transition at the 
scale
of the breaking of the $U(1)_{PQ}$, so that the 
strings 
do not get formed in the first place. This is easily 
achieved. One must worry also about the thermal  
production
of domain walls, but this too is under control 
\cite{ds95,dms95}. 
Of course, unless inflation had taken place before, 
there would be no reason for the Higgs field to have the
same orientation throughout the universe. Thus this 
program
 depends on inflation -here we are completely 
ortodox,
however inflation is allowed to take place at any 
time before the scale of would have been defect 
formation, i.e. 
the $T=0$ mass scale of the theory. This is a general 
feature of all we say in the rest of this talk.

\subsection{Monopole problem}

In order to solve the monopole problem we can either 
nonrestore the grand unified symmetry above the 
unification scale \cite{dms95}or break 
electromagnetic charge 
invariance at nonzero temperatures below it 
\cite{lp80}. Due 
to large next-to-leading terms in gauge theories,
this is hard, if not impossible, to achieve in the 
microscopic scenario. 
It is here where the external background charge of
the universe plays a natural role. In the SM it could 
be a large lepton number, in which case it 
becomes easy to incorporate of electric charge 
\cite{brs97}.
The same can be said of the MSSM. This is in principle
observable (although difficult in practice) and it will
be important to know the content of the neutrino
sea, hopefully to be observed in not too distant 
future.

On the other hand, in supersymmetry the natural 
candidate is provided by the often present global
continous R-symmetries. The MSSM, and its extension
without matter parity,  have an automatic $U(1)$ R-
symmetry
at the renormalizable level, and it is rather easy 
and appealing to construct GUTs with R-symmetries,
as a way of generating the  GUT scale dynamically. 

\subsection{Baryogenesis and spontaneous breaking 
of P and CP}

Why should one resort to the spontaneous breaking of
parity and time-reversal? Well, there is an aesthetic 
motive, for these are fundamental space-time 
symmetries.
More important, spontaneous breaking means less 
divergent
high energy behaviour and this may be instrumental 
in the solution of the strong CP problem. Namely,
in models with spontaneous breaking of these 
symmetries,
especially if supersymmetric \cite{mr96a,mr96b,mrs97},
the strong CP phase can be calculable and small. 
These
theories, though, are plagued with the domain wall 
problem
which can be solved via high T nonrestoration.

The nonrestoration is also crucial for baryogenesis 
\cite{ms80b} in order 
to satisfy the breaking of P and CP as one of the 
three
Sakharov's conditions. This issue was raised in the context 
of $SO(10)$ GUT
\cite{ks80}, where C is a gauge symmetry and is necessarily 
spontaneously broken. If the scale of the breaking of C is 
much below the GUT scale, and if baryogenesis originates at
 the GUT scale, we must have non-restoration \cite{mas82}.

We make one final comment. In the minimal
model of spontaneous CP violation with two Higgs 
doublets
\cite{l73}, in the absence of external charge at 
high 
temperature the symmetry is restored \cite{dms96}. 
Here
the charge (such as  the lepton number of  the SM) makes it 
work
\cite{brs97}, eliminating the domain wall problem and 
making the case for high T baryogenesis.

\section*{Acknowledgments}
 
I wish to acknowledge the local organizers of COSMO97
for their hospitality and a great job in organizing a 
remarkable conference. I am deeply grateful to my 
collaborators Borut Bajc, Gia Dvali, Alejandra Melfo,
Rabi Mohapatra and Antonio Riotto.
This work was in part suported by an EEC grant under the TMR
contract ERBFMRX-CT960090.

\section*{References}

%\bibliographystyle{../biblio/hprsty}
%\bibliography{../biblio/all}

\end{document}